\documentclass[twoside,11pt]{article}

\usepackage[submission]{melba}

\usepackage{graphicx}
\usepackage{pifont}
\usepackage{caption}
\usepackage{subcaption}
\usepackage{dirtytalk}
\usepackage{booktabs}

\usepackage{amsmath,amsfonts}

\melbaheading{0}{https://www.melba-journal.org/article/XX-AA}{2021}{1-?}{mm/yyyy}{mm/yyyy}{Ziller, Usynin, Remerscheid, Knolle, Makowski, Braren, Rueckert, Kaissis}{}{}

\ShortHeadings{DP federated multi-site medical image segmentation}{Ziller, Usynin, Remerscheid, Knolle, Makowski, Braren, Rueckert, Kaissis}
\firstpageno{1}

\title{Differentially private federated deep learning for multi-site medical image segmentation}

\author{\name Alexander Ziller \\ \addr Institute for Artificial Intelligence and Informatics in Medicine, Technical University of Munich, Munich, Germany \\ Institute of Diagnostic and Interventional Radiology, Technical University of Munich, Munich, Germany \\ OpenMined\AND 
\name Dmitrii Usynin \\ \addr Department of Computing, Imperial College London, London, United Kingdom \\ Institute for Artificial Intelligence and Informatics in Medicine, Technical University of Munich, Munich, Germany\\ Institute of Diagnostic and Interventional Radiology, Technical University of Munich, Munich, Germany \\ OpenMined \AND
\name Nicolas W. Remerscheid  \\ \addr Institute for Artificial Intelligence and Informatics in Medicine, Technical University of Munich, Munich, Germany \\OpenMined \AND
\name Moritz Knolle \\ \addr Institute for Artificial Intelligence and Informatics in Medicine, Technical University of Munich, Munich, Germany \\ Institute of Diagnostic and Interventional Radiology, Technical University of Munich, Munich, Germany  \AND
\name Marcus Makowski \\ \addr Institute of Diagnostic and Interventional Radiology, Technical University of Munich, Munich, Germany \AND
\name Rickmer Braren \\ \addr Institute of Diagnostic and Interventional Radiology, Technical University of Munich, Munich, Germany \AND
\name Daniel Rueckert  \\ \addr Institute for Artificial Intelligence and Informatics in Medicine, Technical University of Munich, Munich, Germany \\ Department of Computing, Imperial College London, London, United Kingdom \AND 
\name Georgios Kaissis \\ \addr Institute for Artificial Intelligence and Informatics in Medicine, Technical University of Munich, Munich, Germany \\ Institute of Diagnostic and Interventional Radiology, Technical University of Munich, Munich, Germany \\ Department of Computing, Imperial College London, London, United Kingdom \\ OpenMined
}

\begin{document}

\maketitle

\begin{abstract}
Collaborative machine learning techniques such as federated learning (FL) enable the training of models on effectively larger datasets without data transfer. Recent initiatives have demonstrated that segmentation models trained with FL can achieve performance similar to locally trained models. However, FL is not a fully privacy-preserving technique and privacy-centred attacks can disclose confidential patient data. Thus, supplementing FL with privacy-enhancing technologies (PTs) such as differential privacy (DP) is a requirement for clinical applications in a multi-institutional setting. The application of PTs to FL in medical imaging and the trade-offs between privacy guarantees and model utility, the ramifications on training performance and the susceptibility of the final models to attacks have not yet been conclusively investigated.
Here we demonstrate the first application of differentially private gradient descent-based FL on the task of semantic segmentation in computed tomography. We find that high segmentation performance is possible under strong privacy guarantees with an acceptable training time penalty. We furthermore demonstrate the first successful gradient-based model inversion attack on a semantic segmentation model and show that the application of DP prevents it from divulging sensitive image features. 
Our code is available at~\url{https://github.com/TUM-AIMED/PrivateSegmentation}.
\end{abstract}


\section{Introduction}

Training effective machine learning (ML) models in medical imaging is a data-driven problem, where model utility is typically directly dependent on the quantity and quality of data available during training. In recent works \cite{Sheller2019,Sheller2020}, federated learning (FL) has been proposed to allow the utilisation of multi-site clinical datasets to enable and encourage collaboration between data owners, obtaining larger pools of high quality, diverse and representative data while avoiding direct data sharing. While FL circumvents centralised data pooling, it is not a privacy-enhancing technology (PT), as it does not provide the federation with any formal notion of privacy in regards to the patient data they are holding. This can leave the model vulnerable to catastrophic privacy breaches through attacks such as model inversion \citep{dlg,idlg,geiping2020} or membership inference \citep{mia} during model training by malicious parties inside or outside the federation. It is therefore imperative that collaborative learning does not just benefit the utility of the model through a richer data pool, but also provides formal privacy guarantees to the participating parties, for example through the utilisation of PTs such as differential privacy (DP) or encrypted algorithm training.
However, the application of PTs often comes at the cost of decreased model utility (\textit{privacy-accuracy-trade-off}), e.g. due to the addition of noise to the training process \citep{dp_impact_accuracy}. Minimising this trade-off is a complex, yet fundamental process, and has so far not been conclusively investigated in the area of medical imaging.  \newline
In the present work, we perform federated medical image segmentation under image-level differential privacy guarantees. Through detailed experiments, we demonstrate that the appropriate choice of architecture and training technique can enable excellent model performance despite rigorous privacy guarantees while minimising network overhead and computational demands.

\section{Background and Related Work}
We use the term \textit{Federated Learning }(FL, \citep{konevcny2016federated}) to denote a collaborative learning protocol in which a number of parties (\textit{nodes/workers}) jointly train a neural network. Training occurs in \textit{rounds} during which the \textit{central server} sends the model to the nodes, local training occurs for a number of iterations, whereupon the updated models are aggregated by Federated (Gradient) Averaging \citep{mcmahan2017communication}. We assume a \textit{cross-silo topology}, based on a small number of centres with relatively large datasets. Moreover, we assume homogeneous node compute resources and constant node availability. \newline
We use the following definition of \textit{Differential Privacy} (DP) by \citep{dwork2014algorithmic}:  
For some randomised algorithm (=\textit{mechanism}) \textit{M}, all subsets of its image $S$, sensitive dataset \textit{$D$} and its neighbouring dataset \textit{$D'$}, whereby \textit{$D$} and \textit{$D'$} differ by at most one record, we say that \textit{M} is $(\epsilon, \delta)$-differentially private (DP) if, for a (typically small) constant $\epsilon$ and $0\leq\delta<1$: 
\[Pr(\mathit{M(D) = x}) \leq e ^ \epsilon Pr(\mathit{M(D^{\prime}) = x}) + \delta,  x \in S. \]
The probability $Pr$ is taken over the randomness of the algorithm $M$. DP is, therefore, an attribute of an algorithm that makes it approximately equivariant to exclusion or inclusion of a data point. It quantifies an individuals contribution to the final outcome of the computation. When the difference between the contributions of multiple participants is minimal, it is not possible to reliably determine the presence or the absence of an individual. \par
Sheller and colleagues have demonstrated the utilisation of FL in the context of brain tumour segmentation \citep{Sheller2019,Sheller2020}. However, neither work utilises PTs to provide privacy guarantees to the included patients. Li et al. \citep{li2019privacy} also showcase brain tumour segmentation using FL. The \textit{Sparse Vector Technique} utilised in the study however only provides privacy guarantees to the\textit{ model's parameters} and not to the dataset records, which is not a meaningful notion of privacy. In comparison, DP-SGD, utilised in our study, provides the guarantees to each individual \textit{patient}, providing the federation with an a pragmatic, information-theoretic privacy-preserving solution instead. Fay et al. \citep{fay2020decentralized} utilise the \textit{Private Aggregation of Teacher Ensembles} for brain tumour segmentation. This technique was originally developed for classification tasks and imposes strong assumptions on the learning process. Consequently authors could not demonstrate reasonable privacy guarantees in their study while still experiencing a steep utility penalty. Yang et al. \citep{Yang2021} utilised FL for COVID-19 lesion segmentation in computed tomography (CT) but did not employ PTs. Lastly, Sarma et al. \citep{Sarma2021} demonstrate FL segmentation of prostate volumes in magnetic resonance imaging (MRI), but also did not employ any PTs.

\subsection{Contributions}
As witnessed from the survey of previous works above, although several studies have dealt with the topic of FL for medical image segmentation, our work is, to our best knowledge, the first to utilise differentially private stochastic gradient descent (DP-SGD) in addition to FL in order to provide both stringent image-level privacy guarantees and maintain high model utility in the setting of medical image segmentation. 
We summarise our main contributions below: 
\begin{itemize}
    \item We present an in-depth study on the application of DP to medical image segmentation by successfully training several segmentation model architectures in the federated setting. Contrary to the previous works, our implementation of differentially private training provides strict, provable guarantees with respect to each individual image, which allows the federation to obtain a meaningful measure of privacy.
    \item Our models trained with FL achieve comparable segmentation performance to centrally trained models while suffering only mild privacy-utility trade-offs.
    \item We demonstrate the first successful model inversion attack on semantic segmentation architectures, leading to the full reconstruction of input images for certain models. We thus provide evidence that, despite prevailing literature opinion, FL in itself is an insufficient technique for protecting patient privacy. We then empirically show that -consistent with its theoretical privacy guarantees- the addition of DP to model training completely thwarts such privacy-centred attacks.
\end{itemize}

\section{Methods}
\subsection{Federated Training}
For FL experimentation, we simulated a scenario in which three hospitals (=workers/nodes) are collaboratively training the neural network coordinated by a central server. For this, we split the dataset randomly by patient onto the three servers maintaining an equal number of patients per server. 
We conducted all experimentation using the PriMIA framework \citep{primia}, a generic open-source software package for privacy-preserving and federated deep learning on medical imaging, which we adapted to semantic medical image segmentation. 

\subsection{Differentially Private Training}
For deep neural network training, we utilised DP-stochastic gradient descent (DP-SGD) \citep{abadi2016deep} which extends DP guarantees to gradient-based optimisation by clipping the $L_2$-norm of the per-sample gradients of each minibatch to a specific value and adding Gaussian noise of predetermined magnitude to the averaged minibatch gradients before performing an optimisation step. We utilise the \textit{R\'enyi Differential Privacy Accountant} \citep{mironov2019renyi}, an extension of the \textit{moments accountant} technique by Abadi et al. for the privacy analysis of this algorithm, i.e. the calculation of privacy loss at each individual site in terms of $\epsilon$. We note that the reported privacy guarantees are \textit{record-level}, and not patient-level guarantees. Moreover, we regarded all datasets used as public for the purposes of experimentation such as hyperparameter searches. DP training was performed under three \textit{privacy regimes}, shown in Table \ref{tab:privacy_regimes}. In the following, we will refer to these as \textit{low}, \textit{medium} and \textit{high} privacy regimes. Finally, we note that the utilisation of Batch Normalisation layers is incompatible with DP training, as the running statistics of the layers are maintained non-privately. Hence, we deactivated the running statistics collection for Batch Normalisation layers, effectively converting them to Instance/Channel Normalisation Layers \citep{dai2019channel} which are DP-compatible.

\begin{table}[h]
\centering
\begin{tabular}{@{}lccc@{}}
\toprule
(Hyper-)parameter    & Low                & Medium             & High               \\ \midrule
Noise Multiplier     & $0.8$              & $1.0$              & $1.5$              \\
$L_2$ Clipping Norm  & $1.0$              & $0.5$              & $0.1$              \\
$\delta$ ($\times10^{-5}$) & $1.0$ & $1.0$ & $1.0$ \\
$\epsilon$ local     & $5.98$              & $3.58$              & $1.82$             \\
$\epsilon$ federated & $11.5$             & $7.08$             & $3.54$              \\ \bottomrule
\end{tabular}
\caption{Definition of \textit{privacy regimes} used in the study. The \textit{Noise Multiplier} and \textit{$L_2$ Clipping Norm} are hyperparameters of the DP-SGD algorithm. $\delta$ and $\epsilon$ are \textit{record-level} guarantees for the local model and \textit{per-site record-level} guarantees for the federated models. $\epsilon$ also designates the \textit{privacy budget}, i.e. the privacy loss at which model training is aborted.}
    \label{tab:privacy_regimes}
\end{table}

\subsection{Gradient-based Model Inversion}
In order to empirically verify whether reconstruction of training data by an \textit{Honest-but-Curious} (HbC) adversary with white-box model access \citep{evans2017pragmatic} that participates in the training protocol is possible, we performed a model inversion attack utilising the gradients shared during Federated Averaging. We employ an approach similar to the \textit{Deep Leakage from Gradients} (DLG) \citep{dlg,idlg,geiping2020} attack to infer the data that generated the update. The attack relies on approximating the input image through a gradient descent-based perturbation of a randomly initialised noise matrix, based on a differentiable similarity metric to the gradient captured during model training. Unlike the original attack, designed for an image reconstruction in a classification context given a model update and the corresponding label, we provided the gradient update and a segmentation mask, which are used to reconstruct the input. We observed that the use of a segmentation mask of the victim by the adversary greatly improves the results of the reconstruction. In cases of more complex models, the attack was not successful without access to the segmentation mask, which we consider a limitation of this method.

\subsection{Model Architectures}
All experiments were carried out using U-Net-like architectures \citep{ronneberger2015u} with modifications detailed in \citep{Yakubovskiy2019}, introducing newer architectures as \textit{backbones} to the encoder portion of the U-Net. Since network input/output overhead is a critical bottleneck for federated learning, we focused on architectures which provide a good balance between network size and performance on established benchmarks. Hence, we included MobileNet V2 \citep{sandler2018mobilenetv2} and ResNet-18 \citep{he2016deep} as backbones. Moreover, we considered MoNet \citep{knolle2020efficient}, a novel lightweight U-Net-like architecture with extremely few parameters specifically optimised for FL. Lastly, for comparability to the original U-Net, we utilised an eleven-layer VGG architecture with Batch Normalisation \citep{simonyan2014very} (VGG11 BN). An overview of the number of parameters (i.e. model size) and Multiply/Accumulate operations (MACs) can be found in Table \ref{tab:num_params}. We note that MoNet relies on large receptive field dilated (atrous) convolutions which introduce a substantial number of operations despite the small network size. Moreover, the MobileNet-V2 architecture is optimised for CPU performance \citep{orsic2019defense}. 
We therefore carried out timing experiments both on CPU and GPU (compare Table \ref{tab:timing}).

\begin{table}[h]
    \centering
    \begin{tabular}{@{}lcccc@{}}
    \toprule
         &  MoNet & U-Net  & U-Net  & U-Net  \\
         && (MobileNet V2)&(ResNet-18)& (VGG11 BN)\\\midrule
        Parameters & \textbf{390k} & 6.63M & 14.32M & 18.26M \\
        MACs & 6.11G & \textbf{3.37G} & 5.32G & 14.45G \\
        \bottomrule
    \end{tabular}
    \caption{Number of parameters and Multiply/Accumulate operations for a single forward pass (MACs) of the architectures used in the study. Parameter counts in 32-bit floating point units. \textit{k}: $\times10^3$, \textit{M}: $\times10^6$, \textit{G}: $\times 10^9$. \textbf{Bold} denotes the lowest number of parameters or MACs, correspondingly.}
    \label{tab:num_params}
\end{table}

\subsection{Hyperparameter Optimisation}
We performed hyperparameter optimisation on non-private baseline models in a decentralised manner over the entire federation to obtain suitable values for the learning rate, \textit{beta} parameters for the \textit{Adam} optimiser, as well as translation, rotation and scale values for image augmentation. This corresponds to local hyperparameter optimisation prior to DP training to avoid repeated dataset interaction which would consume the privacy budget. For FL, the synchronisation rate, number of different image augmentations as well as whether or not the model parameters were weighted by the number of samples on the worker during aggregation were additionally optimised. The settings found through optimisation were used to train the corresponding models with DP.

\section{Experiments}
All models were trained and evaluated on the MSD Liver segmentation task \citep{simpson2019large}. The dataset was split into a 63\% training set, 7\% validation set and 30\% held-out testing set. All architectures were pre-trained on a pancreatic CT segmentation dataset from our institution to improve convergence speed and offer equal starting conditions to all models.
\subsection{Segmentation results}
\begin{figure}[h!]
    \centering
    \includegraphics[width=0.9250\textwidth]{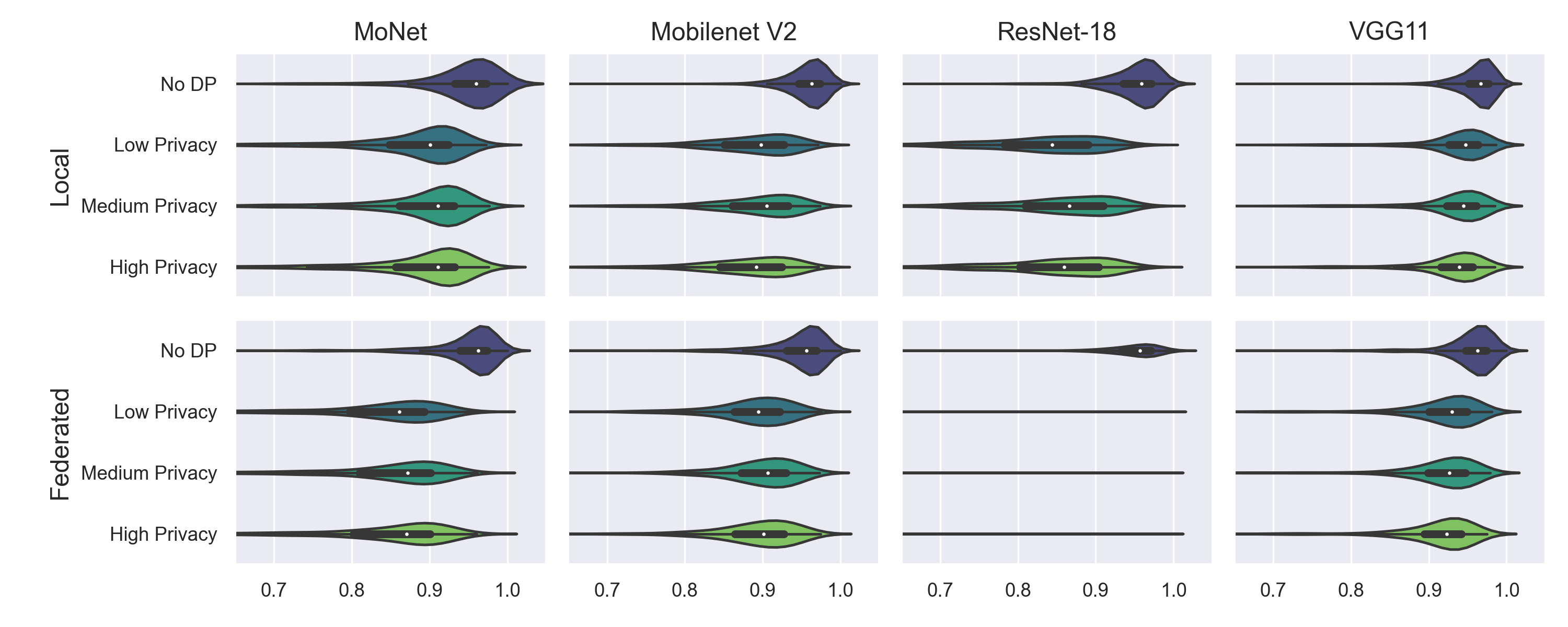}
    \caption{Kernel density estimation plots with inner box plots of Dice scores of the models trained locally (top) and federated (bottom) under different privacy regimes. }
    \label{fig:violins}
\end{figure}

\begin{figure}
    \centering
    \includegraphics[width=0.9250\textwidth]{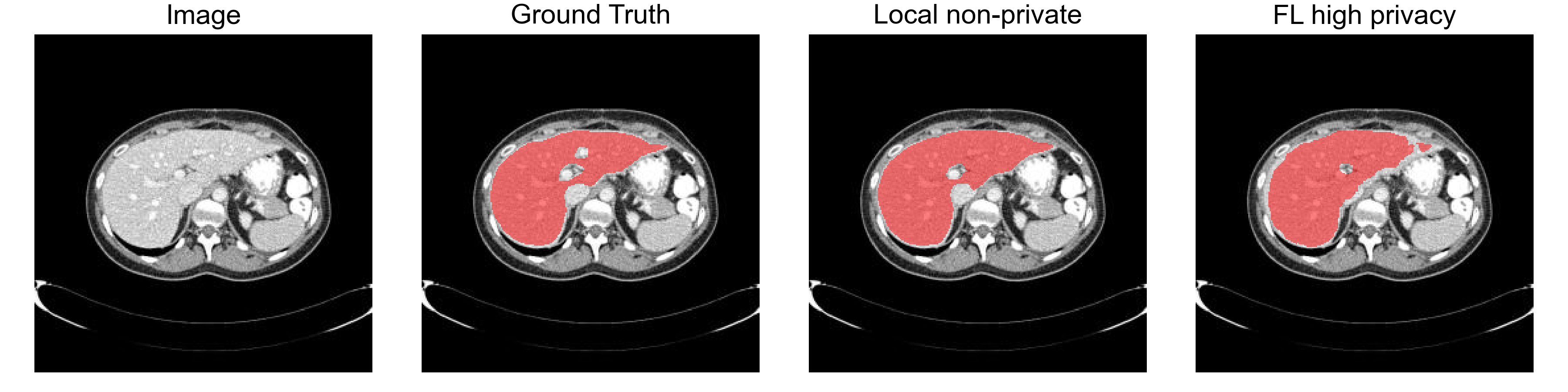}
    \caption{Exemplary segmentation results of the VGG11 BN U-Net architecture, trained in the local non-private setting vs. the FL \textit{high privacy} regime.}
    \label{fig:segmentation}
\end{figure}

Model evaluation results on the test set are listed in Table \ref{tab:scores}. We observed that models trained with FL are able to achieve performance on-par with locally trained models. In line with previous results \citep{abadi2016deep}, all models witnessed performance deterioration due to the utilisation of DP. Surprisingly, we found the performance penalty to be especially high on the ResNet-18 architecture, which did not converge at all in the FL setting with DP. Moreover, we consistently found the VGG-11 BN architecture to perform best. This finding suggests that a high parameter count alone is not the sole determinant of performance deterioration due to DP as suggested in previous work \citep{papernot2019making}. Instead, both the collaborative training setting and the architecture itself seem to contribute to this phenomenon. However, these findings await detailed future analysis. Moreover, the high variance of the per-image Dice score distributions for ResNet-18 and MobileNet V2 backbones suggest a disparate effect of DP on different images (see Figure \ref{fig:violins}), with the MoNet and VGG-11 BN architectures maintaining a more consistent performance in the privately trained local setting. Exemplary segmentation results are shown in Figure \ref{fig:segmentation}.

\begin{table}[h]
\centering
\begin{tabular}{@{}cccccc@{}}
\toprule
Federated & Privacy & MoNet & U-Net  & U-Net  & U-Net \\
&&&(MobileNet V2)&(ResNet-18)& (VGG11 BN)\\ 
\midrule
 & None &93.41&95.41&94.67&\textbf{95.98}\\
 & Low &86.66&87.61&82.20&\textbf{92.85}\\
 & Medium &87.72&88.22&84.36&\textbf{92.63}\\
 & High &87.22&87.04&83.66&\textbf{92.00}\\\hline
\ding{51} & None &94.48&94.13&94.56&\textbf{95.08}\\
\ding{51} & Low &82.10&87.81&\textit{DNC}&\textbf{90.93}\\
\ding{51} & Medium &83.25&88.76&\textit{DNC}&\textbf{90.67}\\
\ding{51} & High &82.75&88.14&\textit{DNC}&\textbf{90.27}\\
\bottomrule 
\end{tabular}
\caption{Dice scores [\%] of models trained locally and with FL under different privacy regimes. We note that ResNet-18 did not achieve convergence under FL and DP (\textit{DNC}). \textbf{Bold} signifies the highest score in the category.}
\label{tab:scores}
\end{table}

In Table \ref{tab:timing}, we compare the time required per epoch of training for the various models. We found the optimised, widely-used architectures to train more efficiently. In comparison, MoNet trades off computational efficiency for a much smaller size to alleviate network input/output constraints. 

 \begin{table}[h]
     \centering
     \begin{tabular}{@{}cccccc@{}}
     \toprule
         Device &Federated &  MoNet & U-Net  & U-Net  & U-Net  \\
          &&& (MobileNet V2)&(ResNet-18)& (VGG11 BN)\\\midrule
           CPU &&$4186.50\pm 13.53$&$769.76\pm 1.68$&$\mathbf{708.50\pm 0.80}$& $959.13\pm 0.43$\\
           CPU & \ding{51} &$4281.64\pm 8.85$&$1200.32\pm 32.96$&$\mathbf{892.62\pm 12.34}$& $1114.57\pm 7.56$\\
          GPU &&$92.84\pm 0.62$ &$43.51\pm 0.32$ &$\mathbf{37.46\pm 0.36}$&$60.73\pm 0.66$\\
          GPU &\ding{51} &$\mathbf{175.11\pm 1.61}$&$458.30\pm 29.66$ &$190.82\pm 5.22$&$192.80\pm 7.59$\\
          \bottomrule
     \end{tabular}
     \caption{Mean and standard deviation [s] for one epoch with equal settings for all models. CPU: Central Processing Unit. GPU: Graphics Processing Unit
     }
     \label{tab:timing}
 \end{table}


\subsection{Model Inversion Attack}
In this study we employ a novel gradient-based model inversion attack based on the works of Geiping et al. \cite{geiping2020}. In our setting the adversary starts with a randomly initialised $<\mbox{\textit{image}}, \mbox{\textit{segmentation  mask}}>$ pair and a captured model update. The optimisation task that is executed by the adversary involves perturbing the image to such extent that it (along with a corresponding segmentation mask) produces a gradient update similar to the one that the adversary captured. We utilise a cosine similarity as the cost function in this optimisation process. The procedure is repeated until either the loss starts diverging or the final iteration is reached. 


We note that for our method the attacker is assumed to have access to a segmentation mask that corresponds to the sensitive image. We consider this to be a limitation of our approach and note that for smaller networks, utilisation of randomly initialised segmentation mask can provide the adversary with a suitable reconstruction result. However, for the purposes of medical imaging, larger models, such as VGG-11 or MoNet require a segmentation mask in order to allow the attacker model to converge and to produce an acceptable reconstruction result.

\begin{figure}[h]
    \centering
    \includegraphics[width=\textwidth]{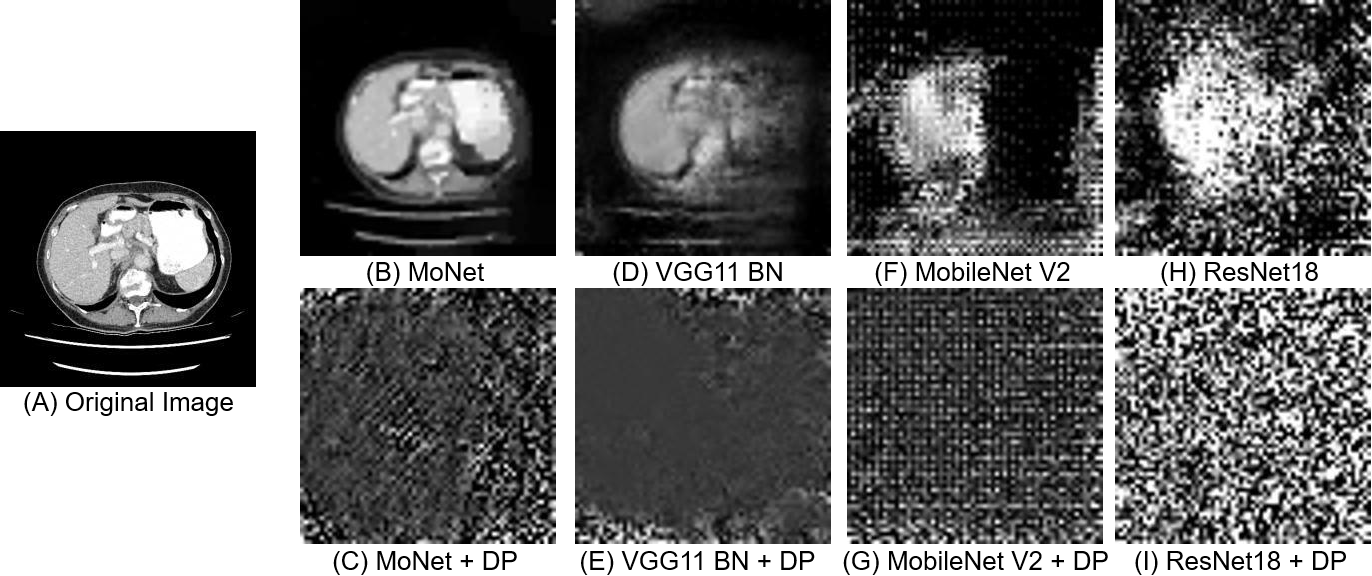}
    \caption{Gradient based reconstructions on the models used in the study. Top row: unprotected models. Bottom row: models with DP (Noise Multiplier 2.0, $L_2$ clipping norm 0.5).}
    \label{fig:attacks}
\end{figure}

Gradient-based model inversion attacks were performed in two settings: initially, gradients captured from FL training without DP were attacked. We then evaluated the attacks on the same architectures with the addition of DP. As presented in Figure \ref{fig:attacks}, non-private models sharing unprotected updates during training risk their data being reconstructed in full. In comparison, models trained with the addition of DP yield no usable information to the inversion attack regardless of architecture. In general, architectural complexity seems to inhibit attack success to a greater degree than shown in previous work on classification \citep{geiping2020}.

\section{Discussion and Conclusion}
To the best of our knowledge, this is the first work to demonstrate DP-SGD-based collaborative model training in the context of semantic medical image segmentation. Our main conclusion is that the provision of rigorous privacy guarantees is possible in the FL setting while maintaining high model utility. One area that we highlight as a promising research direction is an investigation into the relationship between privacy-oriented modes of training and the selection of an optimal model architecture. We found that larger model architectures can be more robust to noise addition, whereas lightweight models can be advantageous in non-private FL settings. This highlights a promising research area that considers the application of privacy-preserving mechanisms in task-specific deployments in order to better tailor defense mechanisms to learning tasks with optimal utility preservation. Additionally, we outline a requirement for investigations into the field of \textit{pragmatic} applications of DP, in order to allow privately trained models to be interpretable by machine learning researchers and facilitate the widespread utilisation of private collaborative model training. Our novel model inversion attack in the unprotected FL setting resulted in a catastrophic privacy breach, while only utilising a segmentation mask and a shared model update. This highlights that FL alone is an insufficient privacy preservation mechanism in collaborative learning and should be regarded as a method for preservation of data ownership/governance which facilitates controlled data \textit{access}. As our method requires possession of a segmentation mask by the adversary, future work will include a natural relaxation of this requirement and the study of robust, scalable model inversion attacks. We note that similarly to other model inversion implementations, supporting large batches of images is a non-trivial task, therefore we also outline this area as a potential future work direction. We expect that our work will stimulate further research on privacy-preserving machine learning, essential to large scale, multi-site medical imaging analysis, in order to allow collaborative model training while mitigating associated privacy risks.

\acks{Georgios Kaissis received funding from the Technical University of Munich, School of Medicine Clinician Scientist Programme (KKF), project reference H14. Dmitrii Usynin received funding from the Technical University of Munich/ Imperial College London Joint Academy for Doctoral Studies. This research was supported by the UK Research and Innovation London Medical Imaging \& Artificial Intelligence Centre for Value Based Healthcare.
The funders played no role in the design of the study, the preparation of the manuscript or the decision to publish. The liver segmentation dataset is described and provided available at \url{https://arxiv.org/pdf/1902.09063}.}

%
\ethics{The work follows appropriate ethical standards in conducting research and writing the manuscript, following all applicable laws and regulations regarding treatment of animals or human subjects.}

\coi{Authors declare no conflicts of interest.}

\bibliography{sample}



{\noindent \em Remainder omitted in this sample. }

\end{document}